%% file: paper.tex
\documentclass[useAMS,usenatbib,usegraphicx]{mn2e}
\usepackage{amssymb,amsmath}
\usepackage{mathrsfs}
\usepackage{times}
\usepackage{setspace}
\usepackage{color}

\include{journal}

\newcommand{\begit}{\begin{itemize}}
\newcommand{\enit}{\end{itemize}}
\newcommand{\begen}{\begin{enumerate}}
\newcommand{\enen}{\end{enumerate}}

\newcommand{\beq}{\begin{equation}}
\newcommand{\eeq}{\end{equation}}
\newcommand{\beqa}{\begin{eqnarray}} 
\newcommand{\eeqa}{\end{eqnarray}} 
\def\lesssim{\mathrel{\hbox{\rlap{\hbox{\lower5pt\hbox{$\sim$}}}\hbox{$<$}}}}
\def\gtrsim{\mathrel{\hbox{\rlap{\hbox{\lower5pt\hbox{$\sim$}}}\hbox{$>$}}}}

\title[Tycho B and SN 1572]{A Quadruple or Triple Origin For Tycho B and SN 1572}

\author[Thompson \& Gould]{Todd A.~Thompson \& Andrew Gould \\
Department of Astronomy and Center for Cosmology \& Astroparticle Physics, 
The Ohio State University, 140 West 18th Avenue, Columbus, OH 43210, USA \\
email: thompson,gould@astronomy.ohio-state.edu}

\begin{document}
\voffset -1.5cm

\maketitle

\label{firstpage}

\begin{abstract}
Kerzendorf et al.~(2012) recently reported the startling discovery of a metal-poor ($[{\rm Fe/H}]\simeq-1\pm0.4$) A-type star near the center of the Tycho supernova remnant.  We propose two possible explanations.  In the first, Tycho B is a blue straggler, formed from the merger of a close K- or G-type binary system, which was previously in a quadruple system with the binary that produced SN 1572.   Both binaries were likely brought to tidal contact by Kozai-Lidov oscillations acting in concert with tidal friction. Analogous progenitor systems may include CzeV343, VW LMi, and KIC 4247791. In the second, Tycho B is the surviving tertiary component of a triple system, which was also likely affected by Kozai-Lidov oscillations.  Rates are briefly discussed.  Problems with each evolutionary scenario are presented. Finally, a chance alignment between Tycho B and the supernova remnant is not excluded.
\end{abstract}

\section{Introduction}
\label{section:introduction}

The progenitors of Type Ia supernovae remain unknown \citep{maoz_mannucci}.  
Recent searches for possible binary companions in their remnants have strongly 
constrained the single-degenerate model \citep{li_2011fe,schaefer}.
The lack of obvious
companions points to the double-degenerate model, wherein
the merger of WD-WD binaries produce Ia supernovae \citep{iben_tutukov,webbink}.

That Ia's from the merger of WD-WD binaries might occur in higher-order multiple 
systems was first discussed by \cite{iben_tutukov_triple}.
\cite{thompson_kozai} showed that a
hierarchical tertiary  
can also affect the evolution of the WD-WD binary through the 
action of Kozai-Lidov cycles \citep{kozai,lidov}, which periodically 
drive the system to high eccentricity, 
strongly decreasing the gravitational wave merger timescale,
as discussed in other contexts
\citep{bls,miller_hamilton}.

Estimates indicate that many Ia supernovae could occur in 
triple systems, motivating searches for the surviving
tertiary in supernova remnants \citep{thompson_kozai}.
However, since the semi-major axis of the tertiary is expected to be 
$\sim1-10^2$\,AU at the  time of the explosion,
it is likely to have normal abundances and kinematics, and to 
have undergone minimal shock heating and ablation \citep{marietta,shappee_overluminous,pan}.  

In addition to these points about the role of multiple systems in 
compact object mergers, the combined effects of Kozai-Lidov oscillations and 
tidal friction likely produce many close binaries \citep{mazeh_shaham,eggleton_kiseleva}.  
\cite{tokovinin} find that the fraction of stellar binaries with tertiary
components strongly increases as the inner binary period decreases, suggesting
that Kozai-Lidov oscillations drive the system to high enough eccentricity 
to reach tidal contact, and that subsequent tidal evolution produces the abundance of short-period
binaries \citep{fabrycky_tremaine}.  The same mechanism has been invoked to explain the production of hot 
Jupiters on $\sim3$\,day orbits \citep{wu_murray}.

\cite{perets_fabrycky} argued that the same mechanism  
could produce blue stragglers by causing stellar mergers, 
leading to stars that otherwise could not exist in the 
old populations of globular clusters \citep{leonard}.  Moreover,
they noted that since the quadruple fraction is comparable to the triple fraction,
there could be double blue straggler binaries, in which mutual Kozai-Lidov oscillations 
brought both binaries to contact and merger.

Finally, it has also been shown that the quadrupole-order expansion of the 
three-body Hamiltonian for the secular evolution of triple systems
does not capture the dynamics when the mass ratio between the 
inner two binary components is large, and the eccentricity of the tertiary is non-zero.
\cite{katz_etal} and \cite{lithwick_naoz} showed that the dynamics
change qualitatively at octupole-order.  In particular,
the inner binary can flip from prograde to retrograde, and vice-versa, 
causing large eccentricity excursions to $1-e\sim10^{-4}-10^{-5}$. \cite{shappee_thompson}
showed that this ``eccentric Kozai mechanism'' (EKM), which is suppressed in nearly 
equal mass binaries, can be triggered by mass loss; after the AGB phase of an intermediate-mass star, 
this could bring a WD to tidal contact with its stellar companion via EKM.  They termed this 
the ``Mass-loss Induced Eccentric Kozai mechanism'' (MIEK).

\cite{kerz}'s discovery of a metal-poor ($[{\rm Fe/H}]\simeq-1\pm0.4$)
rotating ($V \sin i\simeq170\,\,{\rm km\,\,s^{-1}}$)
A-type star (Tycho B) at the center of the Tycho supernova remnant
brings these disparate issues into sharp focus.

\section{Evolutionary Sequences \& Problems}

The first evolutionary sequence we envision is as follows.  A hierarchical quadruple
system forms in the thick disk $\sim10$\,Gyr ago with metallicity $[{\rm Fe/H}]\simeq-1$,
with A/B-A/B ($\simeq 3-5$\,M$_\odot$) and 
K/G-K/G ($0.8-1$\,M$_\odot$) components.  
The initial individual semi-major axes may be $\lesssim1$\,AU,
and the semi-major axis of the orbit of the two systems is $\sim10-100$\,AU.
Our picture is that both 
systems were brought to tidal contact by either normal Kozai-Lidov oscillations 
or a combination of EKM and MIEK, or some yet unknown 4-body mechanism.  
There is also the possibility that the system was simply 
born with two close binaries.  Common envelope and stellar evolution 
in the more massive system  leads to a massive WD-WD binary, which emits gravitational waves 
and evolves to smaller semi-major axis until it eventually merges, producing
SN 1572.   Approximately $0.5-1$\,Gyr before the explosion,  the less massive binary merges,
forming a blue straggler via the mechanism described in \cite{perets_fabrycky},
and producing Tycho B.  

The observed rotation velocity of $V \sin i\simeq170$\,km s$^{-1}$ 
is consistent with intermediate-mass low-metallicity A-type blue 
stragglers \citep{shara}, but it is somewhat higher than is 
observed in old Galactic open clusters \citep{manteiga}. 
Additionally, a rotation velocity of $170$\,km s$^{-1}$ is not exceptional for a field 
A star \citep{zorec_royer}.

The obvious problem with this evolutionary sequence is the kinematics of Tycho B,
which are consistent with a cold thin disk population (Kerzendorf et al.~2012;
their Table 2).  Nevertheless, we see no other way to get a $[{\rm Fe/H}]\simeq-1$
A-type star in the present Milky Way unless it migrated from the outer Galaxy.

The second scenario begins with a thin-disk hierarchical triple system composed of 
an A/B-A/B binary and an A star tertiary born $\sim0.5$\,Gyr ago.  The
massive binary evolves in a way similar to the first evolutionary scenario,
producing SN 1572, and Tycho B is the surviving tertiary.  This scenario has
the problem that it would predict $[{\rm Fe/H}]\sim0$, 
unless the system originated in the far outer Galaxy.

Another possibility is that Tycho B is a chance projection.  This is only 
plausible if  $[{\rm Fe/H}]$ is in fact $\sim0$ since the space density of 
$[{\rm Fe/H}]=-1$ A stars is very small.  Although Tycho B is within $5\arcsec$
of the geometric X-ray center of the remnant from Chandra, \cite{kerz} 
argue that the center is not known to within $30\arcsec$ on the basis of radio data.
The chance alignment to within  $\theta$ at a distance of a few kpc, given a 
space density of $\sim3\times10^{-4}$\,pc$^{-3}$ is $\sim0.01(\theta/5\arcsec)^2$.  
Thus, chance projection is plausible for Tycho B
if the metallicity is Solar, or somewhat sub-Solar.

\section{Quadruple Systems with Close Binaries}

There are a number of quadruple systems known that may consist 
of two close binaries, and which may thus be prototypes for the system 
we envision in the first scenario for SN1572+Tycho-B.  These include 
CzeV343 \citep{cagas}, V994 Her \citep{lee}, OGLE J051343.14-691837.1
\citep{ofir,rivinius}, BD-22 5866 \citep{shkolnik}, VW LMi \citep{pribulla},
KIC 4247791 \citep{lehmann}, and others (e.g., \citealt{pilecki}).

V994 Her is a B8+A0 $+$ A2+A4 quadruple system with 
individual periods of $\simeq2$ and $\simeq1.4$ days, and with an estimated
mutual period of a few thousand years \citep{lee}.  
BD-22 5866 is a K7+K7 $+$ M1+M2 quadruple system with 
individual periods of $\simeq2.2$ and $\leq62$ days, and with an estimated
semi-major axis of $\simeq5.3$\,AU \citep{shkolnik}.  
VW LMi has a contact eclipsing binary with period $\simeq0.48$\,days
with a non-eclipsing detached binary with a period of $\simeq7.9$\,days,
on a mutual orbit with period $\simeq355$\,days and eccentricity $\simeq0.1$.
\cite{pribulla} estimate the total mass of the contact and non-contact binary 
to be 2.4\,M$_\odot$ and 2.2\,M$_\odot$, respectively.
KIC 4247791 is a double eclipsing system 
consisting of a binary of $1.70+1.54$\,M$_\odot$ in a $\simeq4.1$\,day orbit,
together $1.30+1.46$\,M$_\odot$ binary with period $\simeq4.05$\,day.  The 
mutual period is not known, but is constrained to be less than 1200\,yrs 
if the system is in fact a bound quadruple rather than a chance alignment \citep{lehmann}.
Finally, CzeV343  is a double-eclipsing system with inner periods of 
$\simeq1.2$ and $\simeq0.8$ days of roughly A spectral types, respectively,
in a nearly $3:2$ period ratio \citep{cagas}.

Although none of these systems is precisely analogous to that required 
to explain SN 1572+Tycho-B, we consider the existence of such objects to 
essentially prove that the quadruple scenario we propose is possible.
\cite{perets_fabrycky} make the similar point that binary blue straggler
systems should exist since mutual Kozai could produce quadruple systems with 
close binaries that subsequently merge.  Here, we simply appeal to the 
existence of higher mass binaries that might then produce WD-WD binaries 
capable of leading to Ia supernovae.

\section{Rates \& Fractions}

Approximately $2$\%
of all stars born in the mass range $\sim2.5-8$\,M$_\odot$ become Ia supernovae
\citep{horiuchi_beacom}.  If sub-Chandrasekhar mass binaries produce Ia
supernovae, the fraction is smaller.

{\it Triples ---}
About $\sim10$\% of systems are triple \citep{raghavan}, but the uncertainty in this statement 
increases as a function of mass, particularly for systems which might have 
large mass ratios.  In general, we can say that multiplicity is a strong function 
of mass since over 70\% of B- and A-type stars have companions
\citep{shatsky,kouwenhoven}.
For these reasons, \cite{thompson_kozai} argues that in principle all 
Ia supernovae could occur in triple systems, particularly since systems 
composed of, e.g.,  A/B-A/B\,$+$\,G/WD would be very difficult to find.

{\it Quadruples ---}
The fraction of all intermediate-mass stars in quadruple systems consisting of 
two close binaries is more uncertain.
For KIC 4247791, the a priori probability 
for eclipses  of both binaries is $\simeq0.17\times0.1$ if both orbital planes are randomly 
oriented on the sky.  This implies that   
such systems are $\sim60$ times more common than observed.
Since just one system has been found among the $\sim10^4$ intermediate-mass
systems monitored by
Kepler \citep{pinsono}, we conclude that $\sim6\times10^{-3}$ or $\sim0.6$\,\% of 
intermediate-mass systems are hierarchical quadruples consisting of two close  
binaries. Poisson statistics on a single such system suggest that the upper and lower limits on 
the fraction of such systems is $\sim3$ and $\sim10$ times higher and lower, respectively,
than our nominal estimate at 95\% confidence.  

This fraction is consistent with the expectation from multiplicity studies.
The quadruple fraction ($f_{\rm quad}$) is smaller than the triple fraction by $\sim1/3$
\citep{raghavan}.
If we suppose that $\sim0.1$ of these systems are either born close ($f_{\rm close}$), or could
be strongly affected by Kozai-Lidov oscillations, then one expects  
$f_{\rm quad}\times f_{\rm close}\sim0.03\times0.1\sim3\times10^{-3}$ or $\sim0.3$\%
of all systems to be quadruples composed of two close binaries.  
Unfortunately, little is know for massive A/B-A/B binaries in 
quadruple systems with K/G-K/G binaries, as such systems would be
exceedingly difficult to discover and characterize.  

One theoretical argument made
by \cite{shappee_thompson} is that intermediate-mass star binaries
in triple systems (or, presumably, quads)
will in general be more susceptible to MIEK since the ratio of their
ZAMS mass to their WD mass is large, implying that there will be a 
phase after the primary evolves when the mass ratio of the two 
components of the more massive binary is large, leading to the EKM.
For this reason, it may be more likely to produce the type of close binary 
that will lead to Ia supernovae in quads with A/B-A/B components,
and thus the factor $f_{\rm close}$ could be larger than $\sim0.1$
in the above estimate.

{\it Summary---}
Within the very considerable uncertainties, the fraction of 
intermediate-mass triple systems and quadruple systems composed
of two close binaries is consistent with the fraction of 
intermediate-mass stars that become Ia supernovae.

\section*{Acknowledgments}

We thank Ondrej Pejcha, Chris Kochanek,
Marc Pinsonneault, Joe Antognini, Jennifer van Saders,
Jennifer Johnson, and Ben Shappee for discussions. A.G. is supported in part by 
NSF AST-1103471.


\end{document}

%% file: journal.tex
\def\aj{AJ}%
\def\apj{ApJ}%
\def\apjl{ApJ}%
\def\apjs{ApJS}%
\def\aap{A\&A}%
\def\mnras{MNRAS}%
\def\nat{Nature}%


%% file: paper.bbl
\begin{thebibliography}{}

\bibitem[Blaes et al.(2002)]{bls} Blaes, O., Lee, M.~H., \& Socrates, A.\ 2002, \apj, 578, 775 

\bibitem[Caga{\v s} \& Pejcha(2012)]{cagas} Caga{\v s}, P., \& Pejcha, O.\ 2012, \aap, 544, L3 

\bibitem[Eggleton \& Kiseleva-Eggleton(2001)]{eggleton_kiseleva} Eggleton, P.~P., \& Kiseleva-Eggleton, L.\ 2001, \apj, 562, 1012 

\bibitem[Fabrycky \& Tremaine(2007)]{fabrycky_tremaine} Fabrycky, D., \& Tremaine, S.\ 2007, \apj, 669, 1298 

\bibitem[Horiuchi \& Beacom(2010)]{horiuchi_beacom} Horiuchi, S., \& Beacom, J.~F.\ 2010, \apj, 723, 329 

\bibitem[Iben \& Tutukov(1984)]{iben_tutukov} Iben, I., Jr., \& Tutukov, A.~V.\ 1984, \apjs, 54, 335 

\bibitem[Iben \& Tutukov(1999)]{iben_tutukov_triple} Iben, I., Jr., \& Tutukov, A.~V.\ 1999, \apj, 511, 324 

\bibitem[Katz et al.(2011)]{katz_etal} Katz, B., Dong, S., \& Malhotra, R.\ 2011, Physical Review Letters, 107, 181101 

\bibitem[Kerzendorf et al.(2012)]{kerz} Kerzendorf, W.~E., Yong, D., Schmidt, B.~P., et al.\ 2012, arXiv:1210.2713 

\bibitem[Kouwenhoven et al.(2007)]{kouwenhoven} Kouwenhoven, M.~B.~N., Brown, A.~G.~A., Portegies Zwart, S.~F., \& Kaper, L.\ 2007, \aap, 474, 77 

\bibitem[Kozai(1962)]{kozai} Kozai, Y.\ 1962, \aj, 67, 591 

\bibitem[Lee et al.(2008)]{lee} Lee, C.-U., Kim, S.-L., Lee, J.~W., et al.\ 2008, \mnras, 389, 1630 

\bibitem[Lehmann et al.(2012)]{lehmann} Lehmann, H., Zechmeister, M., Dreizler, S., Schuh, S., \& Kanzler, R.\ 2012, \aap, 541, A105 

\bibitem[Leonard \& Linnell(1992)]{leonard} Leonard, P.~J.~T., \& Linnell, A.~P.\ 1992, \aj, 103, 1928 

\bibitem[Li et al.(2011)]{li_2011fe} Li, W., Bloom, J.~S., Podsiadlowski, P., et al.\ 2011, \nat, 480, 348 

\bibitem[Lidov(1962)]{lidov} Lidov, M.~L.\ 1962, Planetary and Space Science, 9, 719 

\bibitem[Lithwick \& Naoz(2011)]{lithwick_naoz} Lithwick, Y., \& Naoz, S.\ 2011, \apj, 742, 94 

\bibitem[Manteiga et al.(1989)]{manteiga} Manteiga, M., Martinez Roger, C., \& Pickles, A.~J.\ 1989, \aap, 210, 66 

\bibitem[Maoz \& Mannucci(2012)]{maoz_mannucci} Maoz, D., \& Mannucci, F.\ 2012, PASA, 29, 447 

\bibitem[Marietta et al.(2000)]{marietta} Marietta, E., Burrows, A., \& Fryxell, B.\ 2000, \apjs, 128, 615 

\bibitem[Mazeh \& Shaham(1979)]{mazeh_shaham} Mazeh, T., \& Shaham, J.\ 1979, \aap, 77, 145 

\bibitem[Miller \& Hamilton(2002)]{miller_hamilton} Miller, M.~C., \& Hamilton, D.~P.\ 2002, \apj, 576, 894 

\bibitem[Naoz et al.(2011)]{naoz_etal} Naoz, S., Farr, W.~M., Lithwick, Y., Rasio, F.~A., \& Teyssandier, J.\ 2011, arXiv:1107.2414 

\bibitem[Ofir(2008)]{ofir} Ofir, A.\ 2008, Information Bulletin on Variable Stars, 5868, 1 

\bibitem[Pan et al.(2012)]{pan} Pan, K.-C., Ricker, P., \& Taam, R.\ 2012, arXiv:1210.0170 

\bibitem[Perets \& Fabrycky(2009)]{perets_fabrycky} Perets, H.~B., \& Fabrycky, D.~C.\ 2009, \apj, 697, 1048 

\bibitem[Pilecki \& Szczygiel(2007)]{pilecki} Pilecki, B., \& Szczygiel, D.~M.\ 2007, Information Bulletin on Variable Stars, 5768, 1 

\bibitem[Pinsonneault et al.(2012)]{pinsono} Pinsonneault, M.~H., An, D., Molenda-{\.Z}akowicz, J., et al.\ 2012, \apjs, 199, 30 

\bibitem[Pribulla et al.(2008)]{pribulla} Pribulla, T., Balu{\v d}ansk{\'y}, D., Dubovsk{\'y}, P., et al.\ 2008, \mnras, 390, 798 

\bibitem[Raghavan et al.(2010)]{raghavan} Raghavan, D., McAlister, H.~A., Henry, T.~J., et al.\ 2010, \apjs, 190, 1 

\bibitem[Rivinius et al.(2011)]{rivinius} Rivinius, T., Mennickent, R.~E., \& Ko{\l}aczkowski, Z.\ 2011, IAU Symposium, 272, 541 

\bibitem[Schaefer \& Pagnotta(2012)]{schaefer} Schaefer, B.~E., \& Pagnotta, A.\ 2012, \nat, 481, 164 

\bibitem[Shappee \& Thompson(2012)]{shappee_thompson} Shappee, B.~J., \& Thompson, T.~A.\ 2012, arXiv:1204.1053 

\bibitem[Shappee et al.(2012)]{shappee_overluminous} Shappee, B.~J., Kochanek, C.~S., \& Stanek, K.~Z.\ 2012, arXiv:1205.5028 

\bibitem[Shara et al.(1997)]{shara} Shara, M.~M., Saffer, R.~A., \& Livio, M.\ 1997, \apjl, 489, L59 

\bibitem[Shatsky \& Tokovinin(2002)]{shatsky} Shatsky, N., \& Tokovinin, A.\ 2002, \aap, 382, 92 

\bibitem[Shkolnik et al.(2008)]{shkolnik} Shkolnik, E., Liu, M.~C., Reid, I.~N., et al.\ 2008, \apj, 682, 1248 

\bibitem[Thompson(2011)]{thompson_kozai} Thompson, T.~A.\ 2011, \apj, 741, 82 

\bibitem[Tokovinin et al.(2006)]{tokovinin} Tokovinin, A., Thomas, S., Sterzik, M., \& Udry, S.\ 2006, \aap, 450, 681 

\bibitem[Webbink(1984)]{webbink} Webbink, R.~F.\ 1984, \apj, 277, 355 

\bibitem[Wu \& Murray(2003)]{wu_murray} Wu, Y., \& Murray, N.\ 2003, \apj, 589, 605 

\bibitem[Zorec \& Royer(2012)]{zorec_royer} Zorec, J., \& Royer, F.\ 2012, \aap, 537, A120 

\end{thebibliography}
